\documentclass[dvipdfm,11pt,a4paper]{article}%{{{
\usepackage{amsmath}
\usepackage{amssymb}
\usepackage{amsthm}
\usepackage{bm}
\usepackage[dvipdfm]{graphicx}
\usepackage[centering,headsep=15pt,head=150pt,foot=35pt,margin=60pt]{geometry}
\usepackage[scaled]{helvet}
\usepackage[slantedGreek]{mathptmx}

\newcommand{\Sla}[1]%
{\kern0.12em{\raise.15ex\hbox{$/$}\kern-.74em #1}}% Feynman slash
\newcommand{\Group}[2]{{ \hbox{{\itshape{#1}}($#2$)} }}
\newcommand{\U}[1]{\Group{U\kern0.05em}{#1}}
\newcommand{\SU}[1]{\Group{SU\kern0.1em}{#1}}
\newcommand{\SL}[1]{\Group{SL\kern0.05em}{#1}}
\newcommand{\Sp}[1]{\Group{Sp\kern0.05em}{#1}}
\newcommand{\SO}[1]{\Group{SO\kern0.1em}{#1}}
\newcommand{\mybar}[1]%
    {{\kern 0.8pt\overline{\kern -0.8pt#1\kern -0.8pt}\kern 0.8pt}}
\newcommand{\sla}[1]%
    {{\raise.15ex\hbox{$/$}\kern-.57em #1}}% Feynman slash
\newcommand{\roughly}[1]%
    {{ \mathrel{\raise.3ex\hbox{ $#1$\kern-.75em\lower1ex\hbox{$\sim$}} } }}
\newcommand{\nop}[1]{:\kern-.3em#1\kern-.3em:}

\newcommand{\del}{\partial}

\newcommand{\delfb}{\overleftrightarrow{\partial}}

\newcommand{\al}{\ensuremath{\alpha}}
\newcommand{\be}{\ensuremath{\beta}}

\newcommand{\de}{\ensuremath{\delta}}

\newcommand{\ep}{\ensuremath{\epsilon}}

\renewcommand{\th}{\ensuremath{\theta}}

\newcommand{\la}{\ensuremath{\lambda}}
\newcommand{\La}{\ensuremath{\Lambda}}

\newcommand{\si}{\ensuremath{\sigma}}
\newcommand{\Si}{\ensuremath{\Sigma}}

\newcommand{\ph}{\ensuremath{\phi}}

\newcommand{\GeV}{ \ensuremath{\mathrm{~GeV}} }
\newcommand{\TeV}{ \ensuremath{\mathrm{~TeV}} }

\newcommand{\hc}{\mathrm{H.c.}} 
\newcommand{\n}{\notag \\}
\newcommand{\mcl}[1]{\mathcal{#1}}

\numberwithin{equation}{section}
\numberwithin{figure}{section}

\begin{document}%{{{
%\pagenumbering{roman}
%\tableofcontents
%\pagestyle{headings}
%\pagenumbering{arabic}
\begin{titlepage}

\begin{flushright}
KEK-TH-1583 \\
UT-12-34
\end{flushright}
\vskip 5em

\begin{center}
{\Large \bfseries
Perturbative unitarity of Higgs derivative interactions \\
}

\vskip 4em

Yohei Kikuta$^\sharp$ \footnote{email: \texttt{kikuta@post.kek.jp}}, \hspace{1em}
Yasuhiro Yamamoto$^\flat$ \footnote{email: \texttt{yamayasu@hep-th.phys.s.u-tokyo.ac.jp}}

\vskip 4em

\begin{center}
$^\sharp$ 
  \textit{KEK Theory Center, KEK,}\\
  \textit{and}\\
  \textit{Department of Particle and Nuclear Physics, The Graduate University for Advanced Studies,}\\
  \textit{1-1 Oho, Tsukuba, Ibaraki 305-0801, Japan}\\[1em]
$^\flat$
  \textit{Department of Physics, Faculty of Science, The University of Tokyo, }\\
  \textit{Tokyo 133-0022, Japan}
\end{center}

\vskip 4em

\textbf{Abstract}
\end{center}

\medskip
\noindent

We study the perturbative unitarity bound given by dimension-six derivative interactions consisting of Higgs doublets.
These interactions emerge from kinetic terms of composite Higgs models or integrating out heavy particles that interact with Higgs doublets.
They lead to new phenomena beyond the Standard Model.

One characteristic contribution from derivative interactions appears in vector boson scattering processes.
Longitudinal modes of massive vector bosons can be regarded as Nambu Goldstone bosons eaten by each vector field.
Since their effects become larger and larger as the collision energy of vector bosons increases, vector boson scattering processes become important in the high-energy region around the TeV scale.
On the other hand, in such a high-energy region, we have to take into account the unitarity of amplitudes.

We have obtained the unitarity condition in terms of the parameter included in the effective Lagrangian for one Higgs doublet models.
Applying it to some models, we have found that contributions of derivative interactions are not large enough to clearly discriminate them from the Standard Model ones.
We also study the unitarity bound in two Higgs doublet models.
Because it is too complex to obtain in the general effective Lagrangian, we have calculated it in explicit models.
These analyses show that the perturbative unitarity bounds are highly model dependent. 

\bigskip
\vfill
\end{titlepage}

\tableofcontents
%}}}
\section{Introduction}%%%%%%%%%%%%%%%%%%%%%%%%%%%%%%%%%%%%%%%%%%%%%%%%%%%%%%%{{{
\label{SecIntro}

Almost all experimental results are consistent with what the Standard Model (SM) predicts.
The recent observations of a new boson \cite{:2012gk} have also been predicted by the model as the Higgs boson.
Measured properties of the observed particle are still consistent with the SM.
However, the model seems to be the effective theory describing physics below the electroweak (EW) scale since it has several theoretical and experimental problems which are maybe explained at the TeV scale. 

One of those problems is how to break the EW gauge symmetry.
Since the symmetry is broken by hand in the SM, the model does not tell us why and how the EW symmetry is broken.
Therefore, we expect the SM should be extended to a model including other sectors responsible for the physics of the electroweak symmetry breaking (EWSB) and phenomena beyond the SM.
If the scale is much higher than the EW scale, the observed Higgs mass requires subtle mechanisms or fine tuning.
Therefore we expect something new to appear in the TeV region.
Eventually, new structure is expected to appear at the TeV scale in order to obtain the EW scale without those artificial constructions.

Many models describing physics beyond the SM have been proposed.
The Higgs sectors of these models are extended, and they break the EW symmetry with a certain mechanism.
The effects of these new sectors could be first observed as deviations from the SM ones via higher-dimensional operators at a scale lower than their original scales.

When we consider physics beyond the SM with extended Higgs sector, its low energy effective theory probably includes dimension-six derivative interactions as a part of the higher-dimensional operators.
These operators have two origins: expansion of kinetic terms if the Higgs doublet is realized as part of a pseudo Nambu Goldstone (NG) field; integrating out heavy new scalar/vector bosons that interact with the Higgs field.
The latter case appears even in models including an elementary Higgs field.

If the Higgs boson were removed from the SM, the Higgs sector would be described by the $SU(2)_L \times SU(2)_R/ SU(2)_V$ nonlinear sigma model.
Derivative interactions of NG fields emerge from the kinetic term. 
These interactions contribute to scattering among longitudinal massive gauge bosons through the equivalence theorem and cross sections of these processes become larger and larger as the energy increases.
They finally become so large as to violate perturbative unitarity around 1 TeV~\cite{Lee:1977eg}.
Of course, the recent observation of the Higgs boson showed us the absence of unitarity violation and the validity of the SM description even much above the TeV scale.
We confront the similar problem in studying derivative interactions of Higgs doublets.
In this paper, we find the scales where the given perturbative description is available in several models that include derivative interactions of the Higgs doublets.

The rest of this paper is organized as follows. 
In Sect.~\ref{SecRev}, we study the unitarity bound given by derivative interactions in one Higgs doublet models (1HDMs).
This is extended to the case of the two Higgs doublet models (2HDMs) in Sect.~\ref{Sec2HDM}. 
In both of these sections, the unitarity violation scales are explicitly calculated with several models.
Finally, our study is concluded in Sect.~\ref{SecConclusion}.

%%}}}
\section{Unitarity of derivative interactions in one Higgs doublet models}%{{{
\label{SecRev}
The perturbative unitarity bounds given by the derivative interaction are discussed on 1HDMs.
First, we derive the formula of the unitarity bound and investigate its general properties.
Then results are applied to explicit models.
The formulae for perturbative unitarity are shown in App.~\ref{AppUni}.

\subsection{Formulae and general properties of the unitarity bound}
The effective Lagrangian of derivative interactions in 1HDM is\footnotemark
\footnotetext{
 Using the field redefinition $H \to H +(a/f^2) (H^\dag H) H$, where $a$ is chosen as an appropriate value, any other dimension-six derivative interaction of the Higgs doublet can be expressed with the kinds of operators given here~\cite{Giudice:2007fh}.
}
\begin{align}
 \mcl{L} \supset&
   \frac{c^H}{2f^2} \del (H^\dag H) \del (H^\dag H)
  +\frac{c^T}{2f^2} (H^\dag \delfb H) (H^\dag \delfb H),
  \label{EqLeffone}
\end{align}
where $f$ is a scale related to new physics and $H^\dag \delfb H := H^\dag (\del H) - (\del H)^\dag H$.
For the second operator, we replace the covariant derivatives with partial ones because in this paper we consider only longitudinal modes of the gauge bosons. 
Since the latter term violates the custodial symmetry, our analysis is based on the Lagrangian with $c^T =0$ \footnotemark.
\footnotetext{
	Since the remaining term, $\del (H^\dag H) \del (H^\dag H)$, changes the normalization of the Higgs field, in addition to the SM contributions, the oblique parameters receive $O(v^2/f^2)$ corrections.
	After eliminating tree-level corrections, models including additional heavy particles probably obtain the oblique corrections at the same order.
	They can be canceled with tuning parameters.
	The derivative interactions receive no sizable corrections from the tuning.
	Therefore, we consider only tree-level contributions to the oblique corrections.
}

Since our focus is entirely on the four point scattering processes given by Eq.~\eqref{EqLeffone}, the vacuum expectation value (VEV) of the Higgs boson plays no role in the following calculation. 
Therefore, we use 
\begin{align}
 H = \begin{pmatrix} C^+ \\ N \end{pmatrix}, \qquad
 H^\dag = \begin{pmatrix} C^-  \ N^\dag \end{pmatrix},
\end{align}
where $C^+ / N$ are a charged/complex-neutral scalar fields.
The charged scalar and imaginary part of the neutral scalar are respectively eaten by $W^\pm$ and $Z$ bosons.
Using the above notation, the following amplitudes are obtained\footnotemark:
\footnotetext{
   We neglect the effects of EWSB because they are below the leading order around the unitarity violation scale.
   For example, the ratio of contributions generated by the SM and the demension-six derivative interactions in the 1HDMs is
\begin{align*}
   \frac{(\text{SM})}{(\text{dim} 6)}
  =\frac{2 m_h^2 f^2}{v^2 c^H \hat{s}}
  \sim \frac{1}{2 c^H} \frac{f^2}{\hat{s}},
\end{align*}
where we have assumed $\hat{s} \gg m_h^2$ and have used the result given by Eq.~\eqref{Eqmatrix1HDM}.
   As we will see later, the typical unitarity violation scale is a few times larger than the decay constant.
   Therefore, the above effects are small enough to be neglected.
}
\begin{align}
 \mcl{M}(C^+ C^- \to C^+ C^-) =& \mcl{M}(N N^\dag \to N N^\dag) \n
 =& \frac{\hat{s}+\hat{t}}{f^2} c^H ,\\
 \mcl{M}(C^+ C^- \to N N^\dag) =&
  \frac{\hat{s}}{f^2} c^H ,
\end{align}
where $\hat{s}$ and $\hat{t}$ are the Mandelstam variables and we consider the energy scale where particles can be treated as massless, i.e., $\hat{s}+\hat{t}+\hat{u}=0$.

Following Ref.~\cite{Lee:1977eg}, we construct matrices with partial wave amplitudes. 
The largest eigenvalue of these matrices gives us the strongest bound to the perturbative unitarity.
We have found that the zeroth mode gives the strongest bound in 1HDMs, so we focus on this case.
With the formulae in App.~\ref{AppUni}, the strongest bound is given by the largest eigenvalue of the following matrix:
\begin{align}
 \begin{pmatrix}
   M_0 (C^+ C^- \to C^+ C^-) & M_0 (C^+ C^- \to N N^\dag ) \\
   M_0 (N N^\dag \to C^+ C^-) & M_0 (N N^\dag \to N N^\dag )
 \end{pmatrix}
 =\frac{\hat{s}}{16\pi f^2}
 \begin{pmatrix}
   c^H/2 &  c^H \\
   c^H   &  c^H/2
 \end{pmatrix}.
 \label{Eqmatrix1HDM}
\end{align}
The perturbative unitarity condition is therefore
\begin{align}
  \frac{\hat{s}}{f^2} \lesssim \frac{16\pi}{3c^H}.
  \label{Equnibound1HDM}
\end{align}
Assuming that derivative interactions are purely given by the kinetic term of the nonlinear sigma model, the conservative cut-off scale is expressed in terms of the decay constant, i.e., $\La \sim 4\pi f.$
\footnote{
  If UV completions are specified, the generalized dimensional analysis may introduce lower cut-off scales~\cite{Georgi:1992dw}.
}
Using the relation, the unitarity bound is related to the cut-off scale as
\begin{align}
  \frac{\hat{s}}{\La^2} \sim \frac{1}{3\pi c^H}.
\end{align}
Therefore, if the relation 
\begin{align}
 c^H \lesssim \frac{1}{3\pi}
\end{align}
is satisfied, models reach the cut-off scale before accessing the unitarity violation scale.
Then, the effective Lagrangian, Eq.~\eqref{EqLeffone}, is available up to the cut-off scale. 
On the other hand, if the coefficient $c^H$ is much larger than unity, the unitarity violation scale is comparable to the scale of the new physics, $f$, so that the description of the effective Lagrangian is invalid even in the energy region around $f$.
In the case $c^H$ is $O(1)$, the unitarity violation scale lies between the new physics scale and the cut-off.
Most of examples shown later are involved in this case.
Around the unitarity bound, we have to include resonance effects; see, for example, Ref~\cite{Contino:2011np}. 
It is therefore necessary to clarify the valid energy scale in the description for each model.

We apply the result to cross sections of the scattering of the Higgs boson and longitudinal modes of massive gauge bosons, so called vector boson scattering (VBS) processes, with the equivalence theorem.
Since these energy scatterings are dominated by the coefficient, $c^H$, with the custodial symmetry, all of the cross sections are proportional to each other.
Here we focus only on the process $W_L^+ W_L^- \to hh$, and relations with the others are shown in Table.~\ref{tbCrossSec1HDM}.
Considering this sort of process, we must remember the importance of the central region\footnotemark
\footnotetext{
  This region is defined as $\cos \th \in [-1/2, 1/2]$ in detectors, where $\th$ is an angle from the beam axis.
}
which is pointed out in Ref.~\cite{Contino:2010mh}, so that we also show the ratios between the cross section of the Higgs pair production and those of the other processes with the central region cut.
The cross section of $W_L^+ W_L^- \to hh$ is
\begin{align}
 \si (W_L^+ W_L^- \to h h) =& 
   \frac{\hat{s}}{32\pi} \left( \frac{c^H}{f^2} \right)^2 \lesssim
	\frac{8\pi}{9\hat{s}} \simeq \frac{1.1 \times 10^6}{\hat{s}\, [\text{TeV}]^2} [\text{fb}] .
\end{align}

\begin{table}[t]
\centering
\begin{tabular}{ccc|rr}
 Process&&& Full & Central \\ \hline
 $W_L^+ W_L^- $&$\to$ &$hh$ & 
   1 & 1/2 \\
 $W_L^+ W_L^- $&$\to$ &$W_L^+ W_L^-$ &
   2/3 & 13/48 \\
 $W_L^+ W_L^- $&$\to$ &$Z_L Z_L$ &
   1 & 1/2 \\
 $Z_L Z_L     $&$\to$ &$hh$ &
   1 & 1/2 \\
 $Z_L Z_L     $&$\to$ &$W_L^+ W_L^-$ &
   2  & 1 \\
 $W_L^+ Z_L   $&$\to$ &$W_L^+ Z_L$ &
   2/3  & 13/48 \\
 $W_L^+ W_L^+ $&$\to$ &$W_L^+ W_L^+$ &
   1 & 1/2
\end{tabular}
\caption{
 Cross-sections of VBS processes in the units of $\si (W_L^+ W_L^- \to hh)$.
 In the Full/Central column, the cross sections of VBS subprocesses with/without the central region cut are shown.
}
\label{tbCrossSec1HDM}
\end{table}

\begin{figure}[t]
\centering
\includegraphics[scale=0.75]{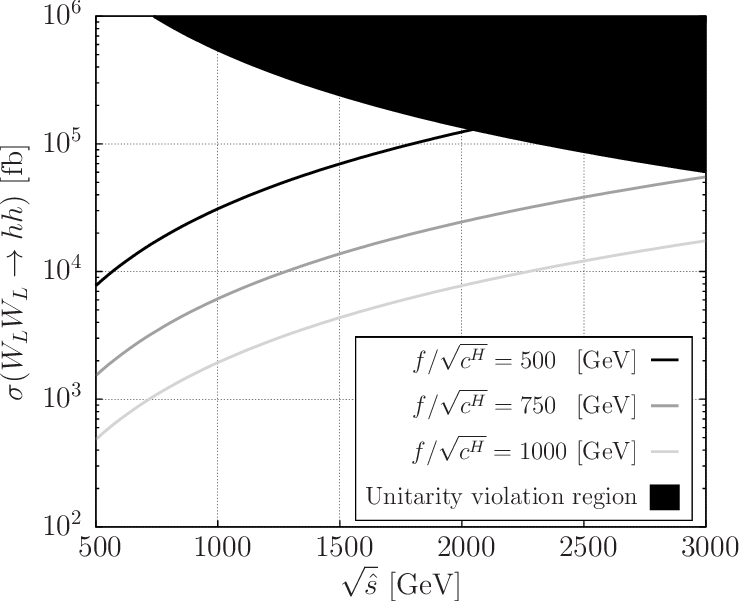}
\caption{
 The upper bound of the cross section for $W_L^+ W_L^- \to hh$ with the perturbative unitarity condition.
 The horizontal axis is the collision energy of this VBS subprocess.
 In the upper shaded region, the unitarity is broken down.
 The black, dark gray and light gray lines are the cross sections where $f/\sqrt{c^H} = 500, 750,$ and $1000$ GeV, respectively.
 For the other processes, the bounds can be obtained with a shift of the vertical axis by the factors given in Table.~\ref{tbCrossSec1HDM}.
}
\label{figWWhh1HDM}
\end{figure}

\noindent
For this process, Fig.~\ref{figWWhh1HDM} shows the region where perturbative unitarity is violated .

Assuming that cross sections reach the above bound at $\sqrt{\hat{s}} = 3\TeV$, we can obtain the relation:
\begin{align}
  \frac{f}{\sqrt{c^H}} \sim \sqrt{\frac{3 \hat{s}}{16 \pi}} \sim 733 \text{ [GeV]}.
\end{align}
If $c^H \sim 1$, the effect of the derivative interaction in the process is comparable to the SM background of about $\sqrt{\hat{s}} = 2 \TeV$, where the cross section is $3 \times 10^4$ fb without the central region cut; see Ref.~\cite{Contino:2010mh}. 
Note that the value of $f$ is typically related to new particle masses.
For example, in the little Higgs scenario~\cite{ArkaniHamed:2001ca}, the top partner mass is given by $O(f)$.
From the viewpoint of the fine tuning, $f$ is required to be below about 1 TeV.

%%}}}
\subsection{Examples with explicit models}%{{{
In the rest of this section, we study the unitarity bounds on two models: the minimal composite Higgs model~\cite{Agashe:2004rs} and the littlest Higgs model with T-parity~\cite{ArkaniHamed:2002qy}.
The latter model has previously been studied in Refs.~\cite{Mahajan:2003vr, Chang:2003vs}.
In Ref.~\cite{Chang:2003vs}, several Little Higgs models are also investigated.
\footnote{
  They obtained the unitarity bounds with all NG bosons.
  However, we focus on the Higgs doublets because other NG bosons are too heavy to treat as massless particles. 
  Our results are conservative compared to theirs.
}
Since the normalization of decay constants can be changed, the combination $f^2 / c^H$ is meaningful.
Here, we follow the normalization given in the original papers.
Decay constants have physical meanings through masses of additional massive vector bosons and fermions in each model.
The Higgs doublet is embedded in such a way as to preserve the custodial symmetry in both models.

%}}}
\subsubsection{The minimal composite Higgs model}%{{{
This model is described by the $SO(5)/SO(4)$ nonlinear sigma model including four NG fields~\cite{Agashe:2004rs}.
They are identified as the Higgs doublet.

The Lagrangian is
\begin{align}
   \mcl{L} =
  \frac{f^2}{2} \left(\del \Si \right)^\dag \left(\del \Si \right) ,
\end{align}
with
\begin{align}
 \Si = \begin{pmatrix}
    \sin [h/f]  \ \vec{h} / h \\
    \cos [h/f]
 \end{pmatrix},
\end{align}
where $\vec{h}$ is the real scalar multiplet of four NG bosons and $h$ is its norm.
Expanding these trigonometric functions, we obtain
\begin{align}
 c^H = 1 .
\end{align}

Using Eq.~\eqref{Equnibound1HDM},  the relation between the decay constant and the energy scale of the unitarity violation is
\begin{align}
 \frac{\hat{s}}{f^2} \sim \frac{16\pi}{3} .
\end{align}
Assuming that perturbative unitarity is violated at $3\TeV$, the decay constant is about $750\GeV$. 
On the other hand, if the decay constant is chosen as $500\GeV$, the perturbativity is preserved up to about $2\TeV$, where the cross section of $W_L^+ W_L^- \to hh$ is $7 \times 10^5$ fb.
In this case, the cross section of the Higgs boson pair production is one order of magnitude larger than that given by the SM.
However, it is challenging to observe this process because the main decay mode is $hh \to 4 b$, which is overwhelmed by the QCD background.

%}}}
\subsubsection{The littlest Higgs model with T-parity}%{{{
Derivative interactions on the littlest Higgs model with T-parity~\cite{ArkaniHamed:2002qy} are shown below.
Scalar fields are described by the $SU(5)/SO(5)$ nonlinear sigma model which includes 14 NG bosons.

The kinetic term of this model is 
\begin{align}
 \mcl{L} =
 \frac{f^2}{8} \text{tr}\left[ 
     \left( \del e^{-2i\Pi/f} \right) \left( \del e^{2i\Pi/f} \right)
  \label{Eqkinetic}
  \right],
\end{align}
where $\Pi$ is the NG field.
The Higgs doublet is assigned in the NG field as
\begin{align}
 \Pi = \frac{1}{\sqrt{2}}
 \begin{pmatrix}
   & H & \\
  H^\dag & & H^T \\
   & H^\ast &
 \end{pmatrix}.
\end{align}
We omitted the other NG bosons since they don't contribute to the current analysis.
Extracting the derivative interaction from the kinetic term, we obtain
\begin{align}
  c^H = \frac{1}{2}.
\end{align}
This result is consistent with previous works~\cite{Mahajan:2003vr, Chang:2003vs}.

If we suppose that $f=750$ GeV, the perturbative unitarity is preserved up to about 4 TeV.
Hence this model description is valid in higher energy scales while the signals of the derivative interaction are smaller than the previous model.
For $f = 750$ GeV, the cross section of $W_L^+ W_L^- \to h h$ in this model almost corresponds to the line where $f/\sqrt{c^H}$ is 1000 GeV in Fig.~\ref{figWWhh1HDM}.

%}}}
\section{Unitarity of derivative interactions on two Higgs doublet models}%{{{
\label{Sec2HDM}

In this section we extend the previous discussion to dimension-six derivative interactions including two Higgs doublets.

The modification is straightforward, and the prescription is also simple.
However, the formulae become too complex because of the many degrees of freedom (DOF).
Then we cannot obtain the formula of the strongest bound like Eq.~\eqref{Equnibound1HDM} with the largest eigenvalue of a matrix that consists of partial wave amplitudes.
Since the matrix can be diagonalized in individual models, three models are investigated as examples.

In this section, we consider processes whose initial states are electromagnetically neutral.
Matrices giving the unitarity bounds for singly or doubly charged initial states are also shown in App.~\ref{AppUnimatrix}.

\subsection{Formulae and general properties of the unitarity bound}

The analyses in this section are based on the following effective Lagrangian:
\begin{align}
 \mcl{L} \supset & 
   \frac{c^H_{1111}}{f^2} O^H_{1111}
  +\frac{c^H_{1112}}{f^2} (O^H_{1112} + O^H_{1121}) \n &
  +\frac{c^H_{1122}}{f^2} O^H_{1122}
  +\frac{c^H_{1221}}{f^2} O^H_{1221}
  +\frac{c^H_{1212}}{f^2} (O^H_{1212} + O^H_{2121})  \n &
  +\frac{c^H_{2221}}{f^2} (O^H_{2221} + O^H_{2212})
  +\frac{c^H_{2222}}{f^2} O^H_{2222} \n &
  +\frac{c^T_{1122}}{f^2} O^T_{1122}
  +\frac{c^T_{1221}}{f^2} O^T_{1221}
  +\frac{c^T_{1212}}{f^2} (O^T_{1212} + O^T_{2121}) ,
  \label{EqLeff}
\end{align}
where
\begin{align}
 O^H_{ijkl} =& 
   \frac{1}{1 +\de_{ik} \de_{jl}} \del (H_i^\dag H_j) \del (H_k^\dag H_l), \\
 O^T_{ijkl} =& 
   \frac{1}{1 +\de_{ik} \de_{jl}} (H_i^\dag \delfb H_j)(H_k^\dag \delfb H_l),
\end{align}
and
\begin{align}
 H_i = \begin{pmatrix} C_i^+ \\ N_i \end{pmatrix}, \qquad
 H_i^\dag = \begin{pmatrix} C_i^- \ N_i^\dag \end{pmatrix}.
\end{align}
In the case where we study the custodial symmetric models, the above coefficients are real and follow the relation derived in App.~\ref{AppCust}:
\begin{align}
  3c^T_{1122} +c^H_{1221} -c^H_{1212} =& 0, \label{Eqcust1} \\
   c^T_{1122} +c^T_{1221} +c^T_{1212} =& 0. \label{Eqcust2}
\end{align}

The 2HDMs require mixing angles to get mass eigenstates of scalar fields.
In this paper, we use the equivalence theorem and focus on only derivative interactions, that is, masses of scalar fields are neglected.
In this case, the perturbative unitarity bound is independent of mixing angles.
This is also true for models including $N$ Higgs doublets.

The unitarity bound is expressed as 
\begin{align}
 \frac{\hat{s}}{f^2} \lesssim \frac{8\pi}{\left| C_\text{max} \right|},
 \label{Equnibound2HDM}
\end{align}
where $C_\text{max}$ is the largest eigenvalue of the matrices given in App.~\ref{AppUnimatrix}.
\footnote{
  Eq.~\eqref{Equnibound2HDM} and the viewpoint explained below are also stated in Ref.~\cite{Chang:2003vs}
}.

As we will see later, the largest eigenvalue $\left| C_\text{max} \right|$ can be as large as about 10.
In this case, the unitarity bound becomes quite strong and leads us to an interesting remark.
Consider, for instance, the pair production of a heavy particle whose mass is $O(f)$ in VBS processes; the energy scale where the pair is produced could be as large as the unitarity violation scale.
This means that we couldn't discuss this kind of process by means of these low-energy descriptions.

%}}}
\subsection{Examples with explicit models}%%%%%%%%%%%%%%%%%%%%%%%%%%%%%%%%%%{{{
\label{Sec2HDM1}

We study the consequences of the above result with several models including two Higgs doublets.
The following three models are studied: the bestest little Higgs model~\cite{Schmaltz:2010ac}; the UV friendly T-parity little Higgs model~\cite{Brown:2010ke}; and an inert doublet model.
The first and second ones are composite Higgs models and the last one is a toy model including elementary Higgs doublets.

%}}}
\subsubsection{The bestest little Higgs model}%{{{
The bestest little Higgs model~\cite{Schmaltz:2010ac} is a little Higgs model which includes two Higgs doublets.
We obtain 15 NG fields that parametrize the $SO(6) \times SO(6) /SO(6)$ coset.
The normalization of the kinetic term is the same as Eq.~\eqref{Eqkinetic}, and the NG field is
\begin{align}
 \Pi = \frac{i}{\sqrt{2}}
 \begin{pmatrix}
   & h_1 & h_2 \\
  -h_1^T & & \\
  -h_2^T & & 
 \end{pmatrix},
\end{align}
where $h_{1,2}$ are real scalar multiplets considered two Higgs doublets and the other NG bosons are eliminated.
In this model, Higgs doublets interact with heavy gauge bosons and a singlet scalar.
The masses of the heavy gauge bosons depend on the other decay constant that is larger than $f$ in order to avoid the constraints from the electroweak precision measurement (EWPM).
Thus the effects coming from the heavy gauge bosons are tiny, and we neglect them.
The interaction with a singlet is required to obtain a collective quartic coupling.
For simplicity, we introduce the following terms to see the effect:
\begin{align}
 \mcl{L}_\si \supset -\frac{m_\si^2}{2} \si^2 + \la f \si (H_1^\dag H_2 +\hc),
\end{align}
where $\si$ is a neutral singlet scalar\footnotemark.
\footnotetext{
 In the original paper~\cite{Schmaltz:2010ac}, $m_\si = \sqrt{\la_{65} +\la_{56}} \, f$ and $\la = \frac{\la_{65} -\la_{56}}{\sqrt{2}}$.
}
Including this contribution, the coefficients of the derivative interactions are
\begin{align}
  c^H_{1111} =& \frac{1}{2}, &
  c^H_{1112} =& 0, &&\\
  c^H_{1122} =& 0, &
  c^H_{1221} =& \frac{1}{4} +c^\si, &
  c^H_{1212} =& \frac{1}{4} +c^\si,\\
  c^H_{2221} =& 0, &
  c^H_{2222} =& \frac{1}{2}, &&\\
  c^T_{1122} =& 0,&
  c^T_{1221} =& \frac{1}{4},&
  c^T_{1212} =&-\frac{1}{4},
\end{align}
where 
\begin{align}
 c^\si = \frac{\la^2 f^4}{m_\si^4}.
\end{align}
The unitarity bound depends on the value of $c^\si$ because the largest eigenvalue is a function of it. 
For $0 \leq c^\si < 1/8$, the bound is
\begin{align}
 \frac{\hat{s}}{f^2} \lesssim \frac{16\pi}{2-c^\si}.
\end{align}
For $c^\si = 0$, it is bounded as 
\begin{align}
 \frac{\hat{s}}{f^2} \lesssim 8\pi,
\end{align}
and it becomes weak as $c^\si$ increases.
For $c^\si =1/8$, the bound is the weakest:
\begin{align}
 \frac{\hat{s}}{f^2} \lesssim 8 \frac{16\pi}{15}.
\end{align}
In the region, $1/8 < c^\si$, the bound is
\begin{align}
 \frac{\hat{s}}{f^2} \lesssim \frac{16\pi}{1 +7c^\si},
\end{align}
where the right-hand side decreases as $c^\si$ increases and the bound becomes the same as the case of $c^\si =0$  at $c^\si = 1/7$.

The unitarity bounds for the cross sections of  $W_L^+ W_L^- \to hh$ and $W_L^+ W_L^+ \to W_L^+ W_L^+$ are displayed below.
We define the mass eigenstates, $h$ and $W_L^\pm$, as follows:
\begin{align}
 \frac{h}{\sqrt{2}} =& N_1^R \cos \al +N_2^R \sin \al ,\\
 W_L^\pm =& C_1^\pm \cos \be +C_2^\pm \sin \be ,
\end{align}
where $N_i^R$ is the real part of $N_i$ and $\al$ and $\be$ are mixing angles.
Unitarity bounds for these processes are
\begin{align}
 \si (W_L^+ W_L^- \to hh) =&
   \frac{\hat{s}}{32\pi f^4} {B_h(\al,\be)}^2
  \lesssim 
  \frac{2\pi}{\hat{s}} \frac{{B_h(\al,\be)}^2}{{C_\text{max}}^2}, \\
 \si (W_L^+ W_L^+ \to W_L^+ W_L^+) =&
   \frac{\hat{s}}{32\pi f^4} {B_w(\be)}^2
  \lesssim
  \frac{2\pi}{\hat{s}} \frac{{B_w(\be)}^2}{{C_\text{max}}^2},
\end{align}
where
\begin{align}
 B_h(\al,\be) =&
   \frac{1}{4}\left( 
	  1 +(1+2c^\si (1-c_{4\be})) c_{2(\al -\be)} +2c^\si s_{4\be} s_{2(\al -\be)}
	\right),\\
 B_w(\be) =&
   \frac{1}{2}\left( 1+c^\si (1 -c_{4\be}) \right).
\end{align}
Here the parameters $c_x$ and $s_x$ are $\cos x$ and $\sin x$,
and $C_\text{max} = (2-c^\si)/2$ for $0\leq c^\si < 1/8$ and $C_\text{max} = (1+7c^\si)/2$ for $1/8\leq c^\si$.
If $\al = \be$ is satisfied, the so-called decoupling limit, we get the relation: $B_h (\be,\be) =B_w (\be)$.

\begin{figure}
\centering
%  \begin{center}
    \begin{tabular}{cc}
    \hspace{-1.35cm}
      {\includegraphics[scale=0.67]{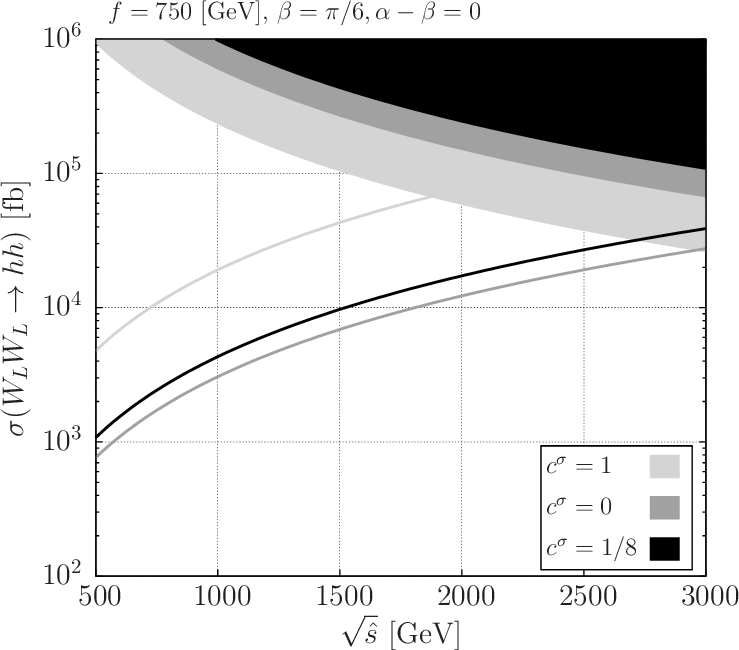}} &
    \hspace{-0.55cm}
      {\includegraphics[scale=0.67]{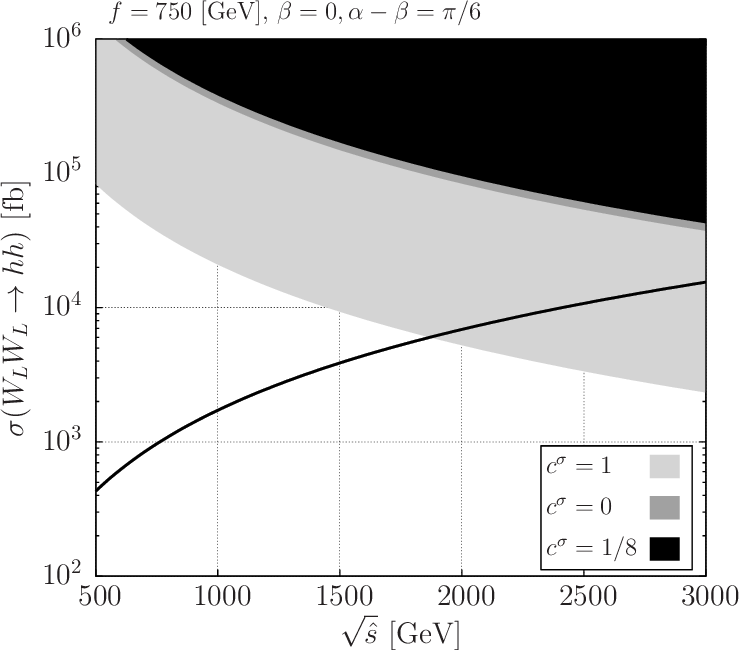}} \\
    \hspace{-1.35cm}  
      {\includegraphics[scale=0.67]{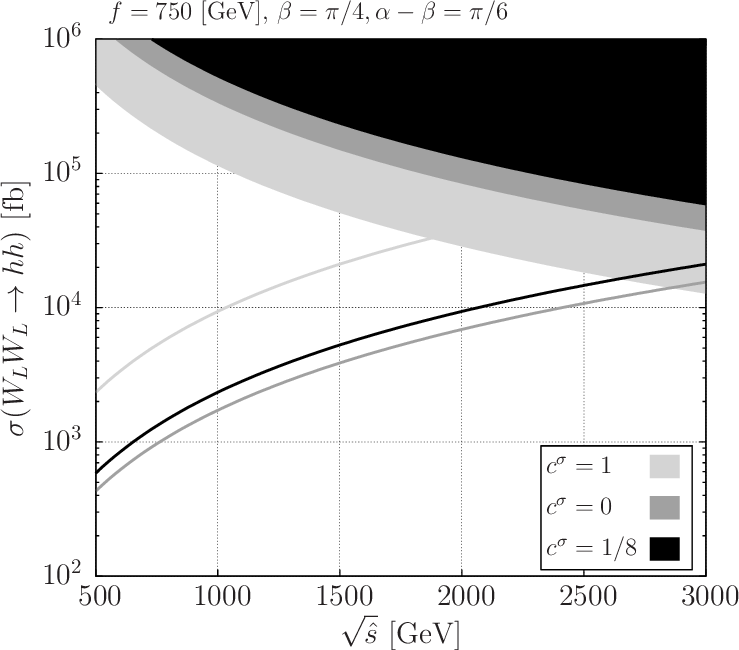}} &
          \hspace{-0.55cm}
      {\includegraphics[scale=0.67]{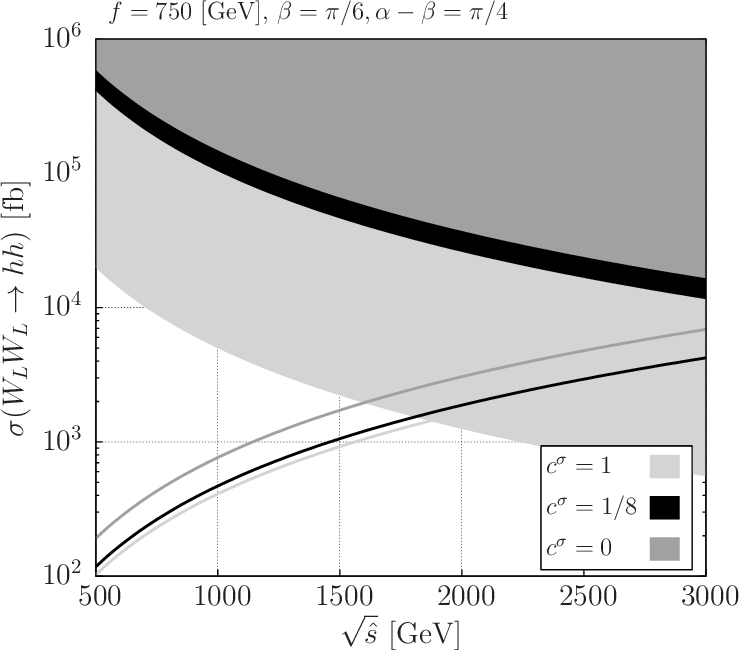}} \\
    \end{tabular}
    \caption{
 The perturbative unitarity bounds of $W_L^+ W_L^- \to hh$ for various $c^\si$ and mixing angles. 
 The decay constant is fixed to be 750 GeV.
 The mixing angles are set to be $(\be , \al - \be) = (\pi / 6 , 0)$ (upper left), $(0 , \pi / 6)$ (upper right), $(\pi / 4 , \pi / 6)$ (lower left) and $(\pi / 6 , \pi / 4)$ (lower right).
 The light gray, dark gray and black lines are cross sections for $c^{\si} = 1,0,$ and $1/8$, respectively.
 The unitarity violation regions depend on the value of $c^\si$, and their brightness corresponds to each line.
    }
    \label{Figbestest}
%  \end{center}
\end{figure}

The perturbative unitarity bounds of $W_L^+ W_L^- \to hh$ are shown in Fig.~\ref{Figbestest}. 
In order to see the effects of the new parameters, we fix the decay constant to be 750 GeV.
The shaded regions in these figures are changed in response to the mixing angles because the cross section depends on the angles.
However, the unitarity bound itself depends only on the coefficient, $c^\si$.
Hence we can see that the energy scales where each cross section line intersects the unitarity violation regions are independent of the angles, e.g., $\sqrt{\hat{s}} \sim 1.9$ TeV for $c^\si = 1$.
For $\be = 0$ and $\al - \be = \pi/6$, the cross sections are independent of the value of $c^\si$; thus, we have only one line but still the intersecting points are the same.

%}}}
\subsubsection{The UV friendly little Higgs model}%{{{
\label{subsec:UVfriendly}
The UV friendly T-parity little Higgs model~\cite{Brown:2010ke} also includes two Higgs doublets as a part of the 14 NG bosons given by the $SU(6)/Sp(6)$ nonlinear sigma model.
The normalization of the kinetic term is also the same as Eq.~\eqref{Eqkinetic}.
This model possesses $Z_2$ symmetry, so-called T-parity, and one of the Higgs doublets is T-odd.
This doublet has no VEV.
Since we study only Higgs doublets, the NG field $\Pi$ can be considered as follows:\footnotemark
\begin{align}
 \Pi = \frac{1}{2}
 \begin{pmatrix}
  & -\ep (H_1 -H_2) & H_1 +H_2 & & \\
  \ep (H_1^\dag -H_2^\dag) & & & -H_1^T -H_2^T \\
  H_1^\dag + H_2^\dag & & & \ep (H_1^T -H_2^T) \\
  & -H_1^\ast -H_2^\ast & -\ep (H_1^\ast -H_2^\ast) &
 \end{pmatrix},
\end{align}
\footnotetext{
 Assignment and normalization of NG bosons given by the original paper are different from the ordinary prescription of the nonlinear sigma model.
 Since we study only the part of two Higgs doublets, normalization of these fields are changed in order to get the canonical kinetic term.
}where $\ep$ is the totally antisymmetric tensor, $\ep^{12} =1$.
Since contributions of heavy new particles can be ignored in this model, derivative interactions are generated only by the kinetic term.
The coefficients of the derivative interactions are as follows:
\begin{align}
  c^H_{1111} =& 4, &
  c^H_{1112} =& 0, && \\
  c^H_{1122} =& 1, &
  c^H_{1221} =& 0, &
  c^H_{1212} =& -3,\\
  c^H_{2221} =& 0, &
  c^H_{2222} =& 4, && \\
  c^T_{1122} =& 1, &
  c^T_{1221} =& 0, &
  c^T_{1212} =& 1.
\end{align}
These coefficients apparently violate the custodial invariant conditions, Eqs.~\eqref{Eqcust1} and \eqref{Eqcust2}.
In this model only one of the Higgs doublets has the VEV, so that tree-level contributions to $\rho$ parameter do not appear.
With these coefficients, the strongest bound is 
\begin{align}
  \frac{\hat{s}}{f^2} \lesssim \pi.
  \label{EquniboundUV}
\end{align}
Assuming that perturbative unitarity is violated at 3 TeV, the decay constant, $f$, is determined as 1.7 TeV.
This value looks large from the viewpoint of fine tuning as we have already seen.
On the other hand, if the decay constant is about 1 TeV, the unitarity is broken below about 1.7 TeV.

\begin{figure}
  \begin{center}
\includegraphics[scale=0.75]{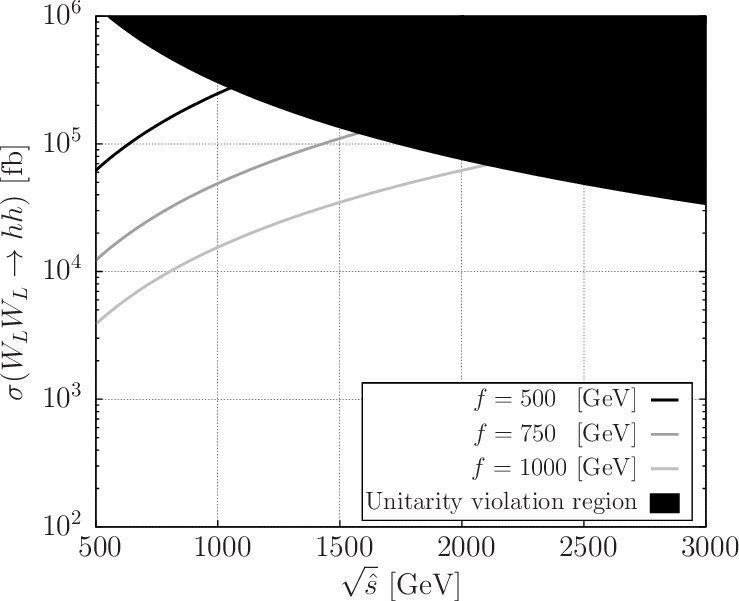}
\caption{
 The perturbative unitarity bounds of $W_L^+ W_L^- \to hh$ in the UV friendly T-parity little Higgs model.
 The horizontal axis is the collision energy of this VBS subprocess.
 In the upper shaded region, unitarity is broken down.
 The black, dark gray and light gray lines are the cross sections corresponding to $f = 500, 750,$ and $1000$ GeV, respectively.
}
\label{WWtohhUV}
  \end{center}
\end{figure}

The unitarity bounds of $W_L^+ W_L^- \to hh$ and $W_L^+ W_L^+ \to W_L^+ W_L^+$ are
\begin{align}
 \si (W_L^+ W_L^- \to hh) =&
   \frac{\hat{s}}{2\pi f^4} 
  \lesssim 
  \frac{\pi}{2\hat{s}} , \\
 \si (W_L^+ W_L^+ \to W_L^+ W_L^+) =&
   \frac{\hat{s}}{2\pi f^4} 
  \lesssim
  \frac{\pi}{2\hat{s}}.
\end{align}
Note that the cross sections have no mixing angle dependence because only one of the Higgs doublet gets a VEV.
These bounds to the cross sections are shown in Fig.~\ref{WWtohhUV}.
They correspond to the case $c^\si = 15/7$ for the bestest little Higgs model.
The unitarity bound of this model is severe because the largest eigenvalue is much larger than the previous models.

%}}}
\subsubsection{Inert doublet models with odd scalars}%{{{
We investigate the following Lagrangian consisting of elementary scalar and vector fields:
\begin{align}
 \mcl{L} \supset &
   -\frac{m_{s0}^2}{2} \ph_0^2 
	 +\la_0 f \ph_0 \left( H_1^\dag H_2 +\hc \right) \n &
	-\frac{m_{sL}^2}{2} \ph_L^a \ph_L^a 
	 +\la_L f \ph_L^a \left( H_1^\dag \si^a H_2 +\hc \right) \n &
	-m_{sL}^2 \ph_T^{a\dag} \ph_T^a 
	 +\sqrt{2}\la_L f \left( \ph_T^{a\dag}(H_1^T \si^2 \si^a H_2) +\hc \right)\n &
	+\frac{m_{v0}^2}{2} V_0 \cdot V_0 
	 +g_0 V_0 \cdot \left( iH_1^\dag \delfb H_2 +\hc \right) \n &
	+m_{v0}^2 V_S^\dag \cdot V_S 
	 +\sqrt{2} g_0 \left( iV_S^\dag \cdot H_1^T \si^2 \delfb H_2 +\hc \right) \n &
   +\frac{m_{vL}^2}{2} V_L^a \cdot V_L^a 
	 +g_L V_L^a \cdot \left( iH_1^\dag \si^a \delfb H_2 +\hc \right).
  \label{EqinertL}
\end{align}
Scalar fields $\ph_0$, $\ph_L^a$, and $\ph_T^a$, are respectively ${\bf 1}_0$, ${\bf 3}_0$, and ${\bf 3}_1$ representations of $SU(2)_L \times U(1)_Y$, and vector fields $V_0$, $V_S$, and $V_L^a$, are, respectively, ${\bf 1}_0$, ${\bf 1}_1$, and ${\bf 3}_0$ representations.
We suppose that these new particles and $H_2$ are odd under an additional $Z_2$ symmetry, and $H_1$ and the other SM particles are even under the discrete symmetry.
We consider the case that only one of the Higgs doublets has a VEV, such as in the model in Sec.~\ref{subsec:UVfriendly}.
These choices of couplings and masses for $\ph_L^a$ and $\ph_T^a$, and $V_0$ and $V_S$ are required to respect $SO(4)$ symmetry.\footnotemark
\footnotetext{
 This structure should be broken by the renormalization group running of the couplings even if the UV completion possesses the structure.
 We assume that this $SO(4)$ symmetry is still good symmetry so as to suppress large corrections to the $\rho$ parameter in the scale where the Lagrangian~\eqref{EqinertL} is available.
}
This set up suppresses contributions to the oblique corrections.

After integrating out heavy particles, we obtain the following coefficients of the derivative interactions:
\begin{align}
 c^H_{1111} =& 0,&
 c^H_{1112} =& 0,&& \\
 c^H_{1122} =& -2s_L +3v_0 +3v_L,&
 c^H_{1221} =& s_0 -2s_L +3v_0,&
 c^H_{1212} =& s_0 -2s_L +3v_L,\label{coeff1_twoH} \\
 c^H_{2221} =& 0,&
 c^H_{2222} =& 0,&& \\
 c^T_{1122} =& -v_0 +v_L,&
 c^T_{1221} =& -s_L -v_L,&
 c^T_{1212} =& s_L +v_0, \label{coeff2_twoH}
\end{align}
where
\begin{align}
 s_0 =& \left( \frac{\la_0 f^2}{m_{s0}^2} \right)^2, &
 s_L =& \left( \frac{\la_L f^2}{m_{sL}^2} \right)^2, \label{Eqcoeff1} \\
 v_0 =& \left( \frac{g_0 f}{m_{v0}} \right)^2, &
 v_L =& \left( \frac{g_L f}{m_{vL}} \right)^2. \label{Eqcoeff2}
\end{align}
Even if additional particles exist, their contributions are included in these four coefficients.
We cannot discriminate these multiple contributions from a large contribution of a particle with a large coupling.
Using these coefficients, the eigenvalues of Eq.~\eqref{Eqneutralmatrix} are fortunately obtained as the following simple forms:
\begin{align}
 c^I_1 =& -\frac{ s_0 -7s_L +3v_0 +3v_L}{2}, \label{Eqeigen1} \\
 c^I_2 =&  \frac{7s_0 -9s_L +9v_0 +9v_L}{2}, \\
 c^I_3 =&  \frac{ s_0 -3s_L +3v_0 -9v_L}{2}, \\
 c^I_4 =&  \frac{ s_0 -3s_L -9v_0 +3v_L}{2}, \\
 c^I_{5\pm} =& \pm \frac{ s_0 + s_L +3v_0 +3v_L}{2}, \label{Eqeigen2} \\
 c^I_{6\pm} =& \pm \frac{ s_0 +9s_L -9v_0 -9v_L}{2}. \label{Eqeigen3}
\end{align}
The strongest unitarity condition is given by Eq.~\eqref{Equnibound2HDM} with the largest eigenvalue in the above.

Derivative interactions generated by integrating out T-odd heavy particles must include two $H_1$ and two $H_2$, as in Eqs.~\eqref{coeff1_twoH} and \eqref{coeff2_twoH}.
Furthermore there are no mixing angles in the Higgs doublets because only one of the Higgs doublets has a VEV.
They are the reason why cross sections of $W_L^+ W_L^- \to hh$ and $W_L^+ W_L^- \to W_L^+ W_L^-$ vanish.

In this model we have four coefficients Eqs.~\eqref{Eqcoeff1} and \eqref{Eqcoeff2} to parametrize the dimension-six differential operators.
If we suppose that $s_0 = s_L = v_0 = v_L =1$, the eigenvalue $c^I_2$ becomes the largest: $c^I_2 =8$.
This value gives us a perturbative unitarity condition which is the same as Eq.~\eqref{EquniboundUV}.

The unitarity bound, Eq.~\eqref{Equnibound2HDM}, can be interpreted as the perturbativity condition of couplings.
For example, if $s_0 = s_L = v_0 = 0$ and $v_L \neq 0$, we get $\left|C_{\text{max}} \right| = 9 v_L / 2$. 
Then the unitarity bound is 
\begin{align}
  \sqrt{\hat{s}} \sim \frac{4 \sqrt{\pi}}{3} \frac{m_{vL}}{g_L}.
\end{align}
In order to preserve unitarity, the unitarity violation scale should be larger than the mass, $m_{vL}$.
As a result, we get the following condition:
\begin{align}
  g_L \lesssim \frac{4 \sqrt{\pi}}{3}.
\end{align}
If the model includes only one $\bf{3}_0$ vector, this requirement is stronger than the naive perturbativity condition $g_L < 4 \pi$.
On the other hand, if it includes several $\bf{3}_0$ vectors, the unitarity bound also limits how many there are.

%}}}
\section{Conclusion}%{{{
\label{SecConclusion}
We have studied perturbative unitarity for dimension-six derivative interactions of the Higgs doublets.
They are generated by kinetic terms in composite Higgs models, or by integrating out heavy particles that interact with the Higgs doublets.
The latter case means that derivative interactions appear even in models consisting of elementary Higgs doublets.

We first studied the unitarity bounds in models including only one Higgs doublet.
The strongest bounds are expressed by the largest eigenvalue of the matrix given by partial wave amplitudes of VBS processes.
We focused on the high-energy region such that derivative interactions could dominate the contributions to the scattering among longitudinal vector bosons.
Assuming that the given derivative interactions respect EWPM, only a combination of parameters, $c^H/f^2$, appears in the analysis.
Therefore, the unitarity condition is expressed by the parameter as Eq.~\eqref{Equnibound1HDM}.
We have applied it to the cross section of $W_L W_L \to h h$.

We have calculated the bounds on explicit models: the minimal composite Higgs model, the littlest Higgs model.
Their structures of global symmetry are significantly different from each other; $SO(5)/SO(4)$ and $SU(5)/SO(5)$.
However, the given bounds are similar; $c^H =1$ and $1/2$.
The decay constants $f$ are related to the masses of the top like fermions in composite Higgs models.
It is therefore supposed that $f/\sqrt{c^H}$ is larger than about 500 GeV;
see e.g., Ref~\cite{:2012uu}, where the perturbative unitarity is violated above the region $\sqrt{\hat{s}} \gtrsim 2\TeV$.
Even in this case, it is difficult to obtain cross sections large enough to distinguish new physics contributions from the SM ones.

Secondly, similar analyses have been performed in 2HDMs.
A simple formula for the unitarity bound could not be obtained in terms of parameters included in the effective Lagrangian~\eqref{EqLeff} since the matrix of partial wave amplitudes is too complex to be diagonalized.
Hence we have investigated the unitarity bound with explicit models: the bestest little Higgs model; the UV friendly T-parity little Higgs model; and the inert doublet model with heavy $Z_2$ odd particles.
The first and the second ones are literally a kind of little Higgs model and the third one is a toy model including elementary Higgs doublets.

In the first one, derivative interactions are generated not only by the kinetic term but also by the integrating out of a heavy scalar field.
The constraints of the former to the unitarity are similar to those given by the 1HDMs discussed in Sec.~\ref{SecRev}.
Including the latter one, the largest eigenvalue depends on the scalar contribution.
The unitarity bound can be stronger than the case not including it.

In the second model, the unitarity condition is much more severe compared to the other models mentioned in this paper.
This is because the coefficients of the derivative interactions are large in this model.
Therefore the unitarity is violated at a low scale compared with the other models.
In this kind of model, large coefficients can produce large cross sections, large enough to exceed the SM background. 
On the other hand, assuming that the masses of additional particles are near the decay constant, they are also near the scale of the unitarity violation.
Therefore, contributions of vector resonances probably need to be considered when people investigate, for instance, pair productions of these additional particles with vector boson collisions.

In the last model that includes an elementary Higgs doublet, VBS processes of the SM particles are suppressed as we have shown.
In this kind of model, the masses of heavy particles should be smaller than the unitarity violation scale.
This condition means that couplings between Higgs doublets and heavy particles are much smaller than the strong coupling, $4\pi$, or the number of these particles is limited.

As a conclusion, we have clarified the importance of studying the unitarity bound when Higgs derivative interactions are investigated because the bound can be significantly lower than the naive cut-off scale.
%}}}
\section*{Acknowledgement}%{{{
We are grateful to Y. Okada for watching over our studies.
Y.Y. would like to thank K. Hamaguchi, T. Moroi and M. Tanabashi for useful comments.
Y.K. would like to thank S. Kanemura for useful discussions.
The work of Y.Y. is supported in part by the Grant-in-Aid for Science Research, Japan Society for the Promotion of Science (JSPS), No. 22244021.
\appendix
%}}}
\section{Perturbative unitarity}%{{{
\label{AppUni}

The amplitudes of elastic scattering satisfy the following relation for each partial wave:
\begin{align}
 M^I_n =& \la (a,b) \left| M^R_n +i M^I_n \right|^2 , \label{EqUnieq} \\
 \la (a,b) =& \sqrt{(1 -(a+b)^2) (1-(a-b)^2)},
\end{align}
where $M^R_n/ M^I_n$ are the real/imaginary parts of the partial wave amplitudes, $a$ and $b$ are the ratios  between the mass of each particle and the center of mass energy, $m_{a,b}/\sqrt{s}$, and partial waves are defined as below with the Legendre polynomials, $P_m(x)$:
\begin{align}
 \mcl{M}(\cos \th) = 16\pi \sum_{n=0}^\infty (2n+1) M_n P_n (\cos \th),\\
 \int_{-1}^1 dx \, P_m(x) P_n(x) =\frac{2}{2n+1} \de_{mn}.
\end{align}
Eq.~\eqref{EqUnieq} is the equation of the circle with radius $1/2\la$ and center $(0,1/2\la)$.
In the high-energy limit where the masses of produced particles can be neglected, the radius of the circle becomes the maximum.
Therefore, the actual amplitudes are in the maximal circle.
Finally, partial wave amplitudes at least satisfy $M^R_n \in [-1/2, 1/2]$ and $M^I_n \in [0, 1]$.
If we consider processes involving identical particles in the final state, the bound becomes weaker as $M^R_n \in [-1, 1]$ and $M^I_n \in [0, 2]$.

Considering the unitarity bound for the derivative interactions, in the massless limit, we can express the amplitudes produced by the derivative interactions as
\begin{align}
 \mcl{M} =&
   \frac{C_s \hat{s} +C_t \hat{t}}{f^2} \\
 =&
   \frac{\hat{s}}{f^2} \left(
	  \left( C_s -\frac{C_t}{2} \right) P_0 +\frac{C_t}{2} P_1
	 \right),
\end{align}
where $\hat{s}$ and $\hat{t}$ are the Mandelstam variables.
The zeroth and the first modes of partial wave amplitudes appear:
\begin{align}
 M_0 =& \frac{\hat{s}}{16\pi f^2} \left( C_s -\frac{C_t}{2} \right), \\ 
 M_1 =& \frac{\hat{s}}{16\pi f^2} \frac{C_t}{6}.
\end{align}
Eventually, the following conditions are respectively obtained, 
\begin{align}
 \frac{\hat{s}}{f^2} \lesssim & \frac{16\pi}{|2C_s -C_t|}, \label{Eqappuni} \\
 \frac{\hat{s}}{f^2} \lesssim & \frac{48\pi}{|C_t|}.
\end{align}
If the condition, $|C_t| \leq 3|C_s| \leq 2|C_t|$, is satisfied, the unitarity bound given by the first mode of the partial wave amplitude is stronger than that given by the zeroth mode.

%}}}
\section{Unitarity matrices and other bounds}%{{{
\label{AppUnimatrix}

Here we derive the matrices given by the zeroth modes of partial wave amplitudes for various VBS processes in 2HDMs.
Using the largest eigenvalue of each, the perturbative unitarity bound is obtained with Eq.~\eqref{Equnibound2HDM}.

\subsection{Neutral two-body states}
The matrix for partial wave amplitudes of neutral two-body states is shown here.
Initial and final states are given by eight states, namely, $C_1^+ C_1^-$, $C_1^+ C_2^-$, $C_2^+ C_1^-$, $C_2^+ C_2^-$, $N_1 N_1^\dag$, $N_1 N_2^\dag$, $N_2 N_1^\dag$ and $N_2 N_2^\dag$.
If all of the coefficients, except for $c^H_{1111}$, are turned off, the matrix becomes the one in the case of 1HDM given in Eq.~\eqref{Eqmatrix1HDM}.
\begin{align}
&
\left(
\begin{matrix}
\frac{3 c^T_{1111}+c^H_{1111}}{2}&
\frac{3 c^T_{1112}+c^H_{1112}}{2}&
\frac{3 c^T_{1112}+c^H_{1112}}{2}&
\frac{3 c^T_{1221}-c^H_{1221}+2 c^H_{1122}}{2} \\
&&&\\
\frac{3 c^T_{1112}+c^H_{1112}}{2}&
\frac{3 c^T_{1122}+2 c^H_{1221}-c^H_{1122}}{2}&
\frac{3 c^T_{1212}+c^H_{1212}}{2}&
\frac{3 c^T_{2221}+c^H_{2221}}{2}\\
&&&\\
\frac{3 c^T_{1112}+c^H_{1112}}{2}&
\frac{3 c^T_{1212}+c^H_{1212}}{2}&
\frac{3 c^T_{1122}+2 c^H_{1221}-c^H_{1122}}{2}&
\frac{3 c^T_{2221}+c^H_{2221}}{2}\\
&&&\\
\frac{3 c^T_{1221}-c^H_{1221}+2 c^H_{1122}}{2}&
\frac{3 c^T_{2221}+c^H_{2221}}{2}&
\frac{3 c^T_{2221}+c^H_{2221}}{2}&
\frac{3 c^T_{2222}+c^H_{2222}}{2}\\
&&&\\
c^H_{1111}&
c^H_{1112}&
c^H_{1112}&
c^H_{1122}\\
&&&\\
c^H_{1112}&
c^H_{1221}&
c^H_{1212}&
c^H_{2221}\\
&&&\\
c^H_{1112}&
c^H_{1212}&
c^H_{1221}&
c^H_{2221}\\
&&&\\
c^H_{1122}&
c^H_{2221}&
c^H_{2221}&
c^H_{2222}
\end{matrix}
\right. \n
& \qquad
\left.
\begin{matrix}
c^H_{1111}&
c^H_{1112}&
c^H_{1112}&
c^H_{1122} \\
&&&\\
c^H_{1112}&
c^H_{1221}&
c^H_{1212}&
c^H_{2221} \\
&&&\\
c^H_{1112}&
c^H_{1212}&
c^H_{1221}&
c^H_{2221} \\
&&&\\
c^H_{1122}&
c^H_{2221}&
c^H_{2221}&
c^H_{2222} \\
&&&\\
\frac{3 c^T_{1111}+c^H_{1111}}{2}&
\frac{3 c^T_{1112}+c^H_{1112}}{2}&
\frac{3 c^T_{1112}+c^H_{1112}}{2}&
\frac{3 c^T_{1221}-c^H_{1221}+2 c^H_{1122}}{2} \\
&&&\\
\frac{3 c^T_{1112}+c^H_{1112}}{2}&
\frac{3 c^T_{1122}+2 c^H_{1221}-c^H_{1122}}{2}&
\frac{3 c^T_{1212}+c^H_{1212}}{2}&
c^T_{2221} \\
&&&\\
\frac{3 c^T_{1112}+c^H_{1112}}{2}&
\frac{3 c^T_{1212}+c^H_{1212}}{2}&
\frac{3 c^T_{1122}+2 c^H_{1221}-c^H_{1122}}{2}&
c^T_{2221} \\
&&&\\
\frac{3 c^T_{1221}-c^H_{1221}+2 c^H_{1122}}{2}&
c^T_{2221}&
c^T_{2221}&
\frac{3 c^T_{2222}+c^H_{2222}}{2}
\end{matrix}
\right).
\label{Eqneutralmatrix}
\end{align}

\noindent
If we impose $SO(4)$ symmetry on the above matrix and eliminate $c^T_{1122}$ and $c^T_{1212}$ with Eqs.~\eqref{Eqcustapp1} and \eqref{Eqcustapp2}, the matrix is simplified as follows:
\begin{align}
&
\left(
\begin{matrix}
\frac{c^H_{1111}}{2}&
\frac{c^H_{1112}}{2}&
\frac{c^H_{1112}}{2}&
\frac{3 c^T_{1221}-c^H_{1221}+2 c^H_{1122}}{2}\\
&&&\\
\frac{c^H_{1112}}{2}&
\frac{c^H_{1221}+c^H_{1212}-c^H_{1122}}{2}&
-\frac{3 c^T_{1221}-c^H_{1221}}{2}&
\frac{c^H_{2221}}{2}\\
&&&\\
\frac{c^H_{1112}}{2}&
-\frac{3 c^T_{1221}-c^H_{1221}}{2}&
\frac{c^H_{1221}+c^H_{1212}-c^H_{1122}}{2}&
\frac{c^H_{2221}}{2}\\
&&&\\
\frac{3 c^T_{1221}-c^H_{1221}+2 c^H_{1122}}{2}&
\frac{c^H_{2221}}{2}&
\frac{c^H_{2221}}{2}&
\frac{c^H_{2222}}{2}\\
&&&\\
c^H_{1111}&
c^H_{1112}&
c^H_{1112}&
c^H_{1122}\\
&&&\\
c^H_{1112}&
c^H_{1221}&
c^H_{1212}&
c^H_{2221}\\
&&&\\
c^H_{1112}&
c^H_{1212}&
c^H_{1221}&
c^H_{2221}\\
&&&\\
c^H_{1122}&
c^H_{2221}&
c^H_{2221}&
c^H_{2222}
\end{matrix}
\right. \n
& \qquad
\left.
\begin{matrix}
c^H_{1111}&
c^H_{1112}&
c^H_{1112}&
c^H_{1122}\\
&&&\\
c^H_{1112}&
c^H_{1221}&
c^H_{1212}&
c^H_{2221}\\
&&&\\
c^H_{1112}&
c^H_{1212}&
c^H_{1221}&
c^H_{2221}\\
&&&\\
c^H_{1122}&
c^H_{2221}&
c^H_{2221}&
c^H_{2222}\\
&&&\\
\frac{c^H_{1111}}{2}&
\frac{c^H_{1112}}{2}&
\frac{c^H_{1112}}{2}&
\frac{3 c^T_{1221}-c^H_{1221}+2 c^H_{1122}}{2}\\
&&&\\
\frac{c^H_{1112}}{2}&
\frac{c^H_{1221}+c^H_{1212}-c^H_{1122}}{2}&
-\frac{3 c^T_{1221}-c^H_{1221}}{2}&
0\\
&&&\\
\frac{c^H_{1112}}{2}&
-\frac{3 c^T_{1221}-c^H_{1221}}{2}&
\frac{c^H_{1221}+c^H_{1212}-c^H_{1122}}{2}&
0\\
&&&\\
\frac{3 c^T_{1221}-c^H_{1221}+2 c^H_{1122}}{2}&
0&
0&
\frac{c^H_{2222}}{2}
\end{matrix}
\right).
\end{align}

\subsection{Singly charged two-body states}
The matrix for the zeroth mode partial wave amplitudes of singly charged two-body states is shown below.
Initial and final states consist of the following four states: $C_1^+ N_1^\dag$, $C_1^+ N_2^\dag$, $C_2^+ N_1^\dag$, and $C_2^+ N_2^\dag$:
% Singly charged without cust
\begin{align}
\begin{pmatrix}
  -\frac{3c^T_{1111}+c^H_{1111}}{2}&
  -\frac{3c^T_{1112}+c^H_{1112}}{2}&
  -\frac{3c^T_{1112}+c^H_{1112}}{2}&
  -\frac{3c^T_{1221}+c^H_{1221}}{2} \\
&&&\\
  -\frac{3c^T_{1112}+c^H_{1112}}{2}&
  -\frac{3c^T_{1122}+c^H_{1122}}{2}&
  -\frac{3c^T_{1212}+c^H_{1212}}{2}&
  -\frac{3c^T_{2221}+c^H_{2221}}{2} \\
&&&\\
  -\frac{3c^T_{1112}+c^H_{1112}}{2}&
  -\frac{3c^T_{1212}+c^H_{1212}}{2}&
  -\frac{3c^T_{1122}+c^H_{1122}}{2}&
  -\frac{3c^T_{2221}+c^H_{2221}}{2} \\
&&&\\
  -\frac{3c^T_{1221}+c^H_{1221}}{2}&
  -\frac{3c^T_{2221}+c^H_{2221}}{2}&
  -\frac{3c^T_{2221}+c^H_{2221}}{2}&
  -\frac{3c^T_{2222}+c^H_{2222}}{2}
\end{pmatrix}.
\end{align}

\noindent
We explicitly impose $SO(4)$ symmetry as in the case of the neutral states:
% Singly charged with cust
\begin{align}
\begin{pmatrix}
  -\frac{ c^H_{1111}}{2}&
  -\frac{ c^H_{1112}}{2}&
  -\frac{ c^H_{1112}}{2}&
  -\frac{3c^T_{1221}+c^H_{1221}}{2} \\
&&&\\
  -\frac{ c^H_{1112}}{2}&
   \frac{ c^H_{1221}-c^H_{1212}-c^H_{1122}}{2}&
   \frac{3c^T_{1221}-c^H_{1221}}{2}&
  -\frac{ c^H_{2221}}{2} \\
&&&\\
  -\frac{ c^H_{1112}}{2}&
   \frac{3c^T_{1221}-c^H_{1221}}{2}&
   \frac{ c^H_{1221}-c^H_{1212}-c^H_{1122}}{2}&
  -\frac{ c^H_{2221}}{2} \\
&&&\\
  -\frac{3c^T_{1221}+c^H_{1221}}{2}&
  -\frac{ c^H_{2221}}{2}&
  -\frac{ c^H_{2221}}{2}&
  -\frac{ c^H_{2222}}{2}
\end{pmatrix}.
\end{align}

\subsection{Doubly charged two-body states}
The entries of the following matrices are the coefficients of VBS processes for the doubly charged states: $C_1^+ C_1^+$, $C_1^+ C_2^+$, and $C_2^+ C_2^+$.
If processes have the same particles in their final states, the unitarity bound becomes weak, as mentioned in App.~\ref{AppUni}.
The effect has been included in the following matrix: 
% Doubly charged without cust
\begin{align}
\begin{pmatrix}
  -\frac{ c^T_{1111}-c^H_{1111}}{2}&
   c^H_{1112}-c^T_{1112}&
  -\frac{ c^T_{1212}- c^H_{1212}}{2}\\
&&&\\
  -\frac{ c^T_{1112}- c^H_{1112}}{2}&
  -\frac{3c^T_{1221}+3c^T_{1122}+c^H_{1221}+c^H_{1122}}{2}&
  -\frac{ c^T_{2221}- c^H_{2221}}{2}\\
&&&\\
  -\frac{ c^T_{1212}- c^H_{1212}}{2}&
   c^H_{2221}-c^T_{2221}&
  -\frac{ c^T_{2222}-c^H_{2222}}{2}
\end{pmatrix}.
\end{align}

\noindent
We also impose $SO(4)$ symmetry on the above:
% Doubly charged with cust
\begin{align}
\begin{pmatrix}
  \frac{ c^H_{1111}}{2}&
         c^H_{1112}&
  \frac{3c^T_{1221}-c^H_{1221}+4c^H_{1212}}{6} \\
&&&\\
  \frac{ c^H_{1112}}{2}&
 -\frac{3c^T_{1221}+c^H_{1212}+c^H_{1122}}{2}&
  \frac{ c^H_{2221}}{2} \\
&&&\\
  \frac{3c^T_{1221}-c^H_{1221}+4c^H_{1212}}{6}&
         c^H_{2221}&
  \frac{ c^H_{2222}}{2}
\end{pmatrix}.
\end{align}

%}}}
\section{Custodial symmetry of derivative interactions}%%%%%%%%%%%%%%%%%%%%%{{{
\label{AppCust}
Dimension-six derivative interactions of Higgs doublets naively violate custodial symmetry.
We study the conditions of the derivative interactions that ensure symmetry in the case of 2HDMs.
In the following discussion, we refer to results and notations in Ref.~\cite{Kikuta:2011ew}.

Derivative interactions are classified into three kinds of operators: operators including unique indices are called type I, e.g. $\del(H_1^\dag H_1) \del(H_1^\dag H_1)$; in type II, only one of four doublets has a different index, e.g. $\del(H_1^\dag H_1) \del(H_1^\dag H_2)$; the others belong to type III, e.g. $\del(H_1^\dag H_2) \del(H_2^\dag H_1)$.
For type I and II, current-current interactions, namely $O^T$ operators, violate custodial symmetry because they produce additional contributions to the mass of the $Z$ boson. 
This is interpreted from a different viewpoint with the nonlinear representation.
For type I, as studied in Ref.~\cite{Low:2009di}, operators belonging to $O^T$ are built from the generator of the hypercharge, i.e., the third generator of $SU(2)_R$:
\begin{align}
  (h T^{R3} \partial h)(h T^{R3} \partial h) = \frac{1}{2} O^T_{1111},
\end{align}
where $h$ is a real scalar multiplet that corresponds to the Higgs doublet.
Since the generator violates $SO(4)$ symmetry, custodial symmetry cannot be preserved after the EWSB.
In other words, operators that consist of $SO(4)$ symmetric combinations of generators produce the custodial symmetric derivative interactions.
This is also the case for operators of type II. \footnotemark
\footnotetext{
 It is also true for imaginary DOF of derivative interactions.
 Any imaginary DOF of coefficients violate, in addition to CP symmetry, custodial symmetry.
 There are no relations to ensure custodial symmetry for imaginary DOF, so we discuss only real DOF here.}

For type III derivative interactions, the situation is different, that is, certain combinations of $SU(2)_R$ violating operators recover $SU(2)_R$ symmetry. 
In this type, the following operators produce real DOF of derivative interactions:
\begin{align}
 T_{11}^{L\al} T_{22}^{L\al} =& 
  \frac{1}{4}( 3O^H_{1221} -O^T_{1122} +O^T_{1221} ), \\
 T_{12}^{L\al} T_{12}^{L\al} =& 
  \frac{1}{2} ( 3O^H_{1122} +3O^H_{1212} +3O^H_{2121} +O^T_{1122} -O^T_{1221} ), \\
 T_{11}^{R\be} T_{22}^{R\be} =& 
  \frac{1}{4} ( 3O^H_{1212} +3O^H_{2121} +O^T_{1122} -O^T_{1212} -O^T_{2121} ), 
   \label{EqTriijj} \\
 T_{12}^{R\be} T_{12}^{R\be} =& 
  \frac{1}{2} ( 3O^H_{1122} +3O^H_{1221} -O^T_{1122} +O^T_{1212} +O^T_{2121} ), 
   \label{EqTrijij} \\
 S_{12}^{\al 3} S_{12}^{\al 3} =& 
  \frac{1}{2} ( 3O^H_{1122} -3O^H_{1212} -3O^H_{2121} +O^T_{1122} -O^T_{1221} ),
   \label{EqS3ijij} \\
 S_{12}^{\al \be} S_{12}^{\al \be} =& 
  \frac{3}{2} ( 2 O^H_{1122} - O^H_{1221} - O^H_{1212} - O^H_{2121}), 
   \label{EqSijij} \\
 T_{11}^{R3} T_{22}^{R3} =& 
  \frac{1}{2} O^T_{1122},
   \label{EqTr3iijj} \\
 T_{12}^{R3} T_{12}^{R3} =& 
  \frac{1}{2} ( O^T_{1221} +O^T_{1212} +O^T_{2121} ), 
   \label{EqTr3ijij} \\
 U_{12} U_{12} =& \frac{1}{2} ( O^T_{1221} -O^T_{1212} -O^T_{2121} ),
\end{align}
where $X_{ij}^A := (h X_{(i,j)}^A \del h)$ and $h$ includes eight real scalar fields interpreted as two Higgs doublets.
Generators $T^{L\al}_{(i,j)}$, $T^{R\be}_{(i,j)}$, $S^{\al\be}_{(i,j)}$ and $U_{(i,j)}$ are, respectively, $({\bf 3}, {\bf 1})$, $({\bf 1}, {\bf 3})$, $({\bf 3}, {\bf 3})$, and $({\bf 1}, {\bf 1})$ representations of $SU(2)_L \times SU(2)_R$.
Explicit forms of these matrices are given in Ref.~\cite{Kikuta:2011ew}.
If we naively follow the discussion for type I and II, the coefficients of operators \eqref{EqS3ijij}, \eqref{EqTr3iijj}, and \eqref{EqTr3ijij} should vanish for $SO(4)$ symmetry.
However, we have found that some operators preserving $SU(2)_R$ can be given by a certain combination of operators violating $SU(2)_R$.
Since these operators are not linearly independent of each other, several relations are derived.
In these relations, the above operators violating $SU(2)_R$ symmetry appear with a certain proportional relation:
\begin{align}
 a_{1212}^S : a_{1122}^Y :a_{1212}^Y = 1:-2:1 ,
\end{align}
where they are respectively coefficients of operators \eqref{EqS3ijij}, \eqref{EqTr3iijj} and \eqref{EqTr3ijij}.
This condition is expressed as
\begin{align}
 c^T_{1122} + c^T_{1221} +c^T_{1212} =& 0,  \label{Eqcustapp1} \\
 3c^T_{1122} -c^H_{1212} +c^H_{1221} =&0 \label{Eqcustapp2} ,
\end{align}
where $c_{ijkl}^{H,T}$ are defined as the coefficients of $O^{H,T}_{ijkl}$.
The result is consistent with the custodial symmetric conditions shown in Refs.~\cite{Kikuta:2011ew}.

The above analysis is easily extended to models including $N$ Higgs doublets.
In this case, two other classes of derivative interactions should be defined: operators including three deferent indices are called type IV, e.g. $\del (H_i^\dag H_j) \del (H_i^\dag H_k)$; the other operators whose indices are totally different from each other are classified as type V, e.g. $\del (H_i^\dag H_j) \del (H_k^\dag H_l)$.
With similar discussions to those given above, the following proportional relations are obtained:
\begin{align}
  a^S_{ijik} : a^Y_{iijk} : a^Y_{ijik} &= 1 : -2 : 1, \\
  a^S_{ijkl} : a^S_{ijkl} : a^Y_{ijkl} : a^Y_{ikjl} : a^Y_{iljk} &= 
    1 : -1 : 1 : -2 : 1, \\
  a^S_{ikjl} : a^S_{iljk} : a^Y_{ijkl} : a^Y_{ikjl} : a^Y_{iljk} &=
    1 : 1 : -1 : 1 : 1 ,
\end{align}
where $a^Y_{ijkl}$ and $a^S_{ijkl}$ are, respectively, coefficients of $T^{R3}_{ij} T^{R3}_{kl}$ and $S^{\al 3}_{ij}S^{\al 3}_{kl}$, the first relation is for type IV and the others are for type V.
The following relations are induced for the coefficients of the derivative interactions:
for type IV,
\begin{align}
 c^T_{iijk} + c^T_{ijik} +c^T_{ijki} =& 0, \\
 3c^T_{iijk} +c^H_{ijki} -c^H_{ijik} =&0 ;
\end{align}
for type V,
\begin{align}
 c^H_{ijkl} -c^H_{ijlk} -c^H_{ikjl} +c^H_{iklj} +c^H_{iljk} -c^H_{ilkj} =& 0, \\
 3(c^T_{ijkl} +c^T_{ijlk}) -c^H_{ikjl} +c^H_{iklj} -c^H_{iljk} +c^H_{ilkj} =& 0, \\
 3(c^T_{ikjl} +c^T_{iklj}) -c^H_{ijkl} +c^H_{ijlk} +c^H_{iljk} -c^H_{ilkj} =& 0, \\
 3(c^T_{iljk} +c^T_{ilkj}) +c^H_{ijkl} -c^H_{ijlk} +c^H_{ikjl} -c^H_{iklj} =& 0.
\end{align}
After imposing these conditions to ensure $SO(4)$ symmetry on the derivative interactions, we get their remaining DOF.
The result is shown in Table.~\ref{TabDOF}.

\begin{table}[t]
\centering
\begin{tabular}{r|ll}
      & with                 & without              \\ \hline
 I    & $N$                  & $2N$                 \\
 II   & $N(N-1)$             & $2N(N-1)$            \\
 III  & $2N(N-1)$             & $3N(N-1)$            \\
 IV   & $2N(N-1)(N-2)$        & $3N(N-1)(N-2)$       \\
 V    & $N(N-1)(N-2)(N-3)/3$ & $N(N-1)(N-2)(N-3)/2$ \\ \hline
 Sum  & $N^2(N^2+2)/3$       & $N^2(N^2+3)/2$       
\end{tabular}
\caption{
Real DOF of dimension-six derivative interactions on models including $N$ Higgs doublets with/without $SO(4)$ symmetry for each type.
}
\label{TabDOF}
\end{table}
%}}}
\bibliographystyle{JHEP} %{{{

%}}}
\end{document}